\documentclass[]{raa}            
\usepackage{graphicx,times}
\usepackage{natbib}

\providecommand{\bm}[1]{\mbox{\boldmath$#1$\unboldmath}}

\begin{document}

   \title{A Binary Study of Colour-Magnitude Diagrams of 12
   Globular Clusters
}

 \volnopage{ {\bf 2009} Vol.\ {\bf 9} No. {\bf XX}, 000--000}
   \setcounter{page}{1}

   \author{Zhong-Mu Li
         \inst{1,2}
   \and Cai-Yan Mao
      \inst{1}
   \and Ru-Heng Li
      \inst{1}
   \and Ru-Xi Li
      \inst{1}
   \and Mao-Cai Li
      \inst{1}
   }

   \institute{College of Physics and Electronic Information, Dali University,
             Dali 671003, China; {\it zhongmu.li@gmail.com}\\
                    \and
             National Astronomical Observatories, the Chinese Academy of
Sciences, Beijing, 100012, China\\
\vs \no
   {\small Received [year] [month] [day]; accepted [year] [month] [day] }
}

\abstract{Binary stars are common in star clusters and galaxies, but
the detailed effects of binary evolution are not taken into account
in some colour-magnitude diagram (CMD) studies. This paper studies
the CMDs of twelve globular clusters via binary-star stellar
populations. The observational CMDs of star clusters are compared to
those of binary-star populations, and then the stellar
metallicities, ages, distances and reddening values of these star
clusters are obtained. The paper also tests the different effects of
binary and single stars on CMD studies. It is shown that binaries
can fit the observational CMDs of the sample globular clusters
better compared to single stars. This suggests that the effects of
binary evolution should be considered when modeling the CMDs and
stellar populations of star clusters and galaxies.
 }

   \authorrunning{Z.-M. Li et al.}            
   \titlerunning{Binary Study of Colour-Magnitude Diagrams of Globular Clusters}  
   \maketitle


%
%
\section{Introduction}           
\label{sect:intro}

Star cluster is a kind of important object in astrophysical studies.
The colour-magnitude diagrams (CMDs) are usually used to determine
the properties of star clusters, i.e., stellar age, metallcity,
distance, and reddening values (see, e.g., \citealt{naylor06}, and
\citealt{kalirai04}). In most works, the theoretical isochrones of
some single-star simple stellar populations (ssSSPs) (e.g.,
\citealt{momany03}) are used , because their corresponding CMDs are
shown as simple curves and are convenient for studies. Meanwhile,
some works use observed template CMDs (\citealt{recio-blanco05}) for
similar studies. Recently, some works have tried to study the CMDs
of star clusters via more advanced techniques, i.e., Monte Carlo
simulations. One can refer to the papers, e.g., \cite{kalirai04},
\cite{aparicio90}, \cite{tosi91}, \cite{hurley98},
\cite{skillman02}, for more information. Such methods are more
advanced and can give additional parameters such as binary fraction
and star formation histories to star clusters. This is an important
progress because these methods take the effects of binaries into
account. Although the new methods are much more informative and
rewarding, they usually need more constraints on the properties of
star clusters from independent techniques to make the fitting
exercisable, and such methods are used to study some star clusters
with accurate observed CMDs. Therefore, it is difficult to study the
CMDs of a big sample of star clusters as the limitation of the
accuracy of observed CMDs and known properties of star clusters. In
addition, most works focus on the widening of the main sequences of
theoretical CMDs when considering the binary effects in the
comparisons of observational and theoretical CMDs (e.g.,
\citealt{naylor06}). However, according to the works of
\cite{kalirai04} and \cite{li08}, binary evolution affects not only
the shapes of some special parts of CMDs, but also the number
distributions of stars in CMDs. Therefore, it is necessary to take
the detailed effects of the evolution of binaries in CMD studies.

According to the studies of \cite{li08} and \cite{li09}, binary
evolution can affect the integrated features of populations. Besides
the widening of isochrones, binary evolution can explain some
special stars such as blue stragglers in star clusters. They can
also interpret the UV excess in elliptical galaxies (see, e.g.,
\citealt{han05}). Therefore, it is calling for CMD studies on the
basis of some stellar populations that have taken the detailed
effects of binary evolution into account. If the effects of binary
evolution are taken into account, some different properties of star
clusters may be obtained. This is helpful for understanding star
clusters further. Furthermore, because stellar isochrone is a basic
ingredient to build stellar populations, to compare the CMDs of star
clusters with the synthetic CMDs of binaries can help us to model
stellar populations in more advanced ways. This paper aims to study
the CMDs of a few star clusters via binary stars, and then compare
the results determined by binary- and single-star simple stellar
populations (bsSSPs and ssSSPs). Because globular clusters (GCs) are
usually thought to be simple systems that include metal-poor and old
stars, we take some GCs as the star cluster sample of this study.

The paper is organized as follows. In Section 2, we introduce our
sample GCs and their observational CMDs. In Section 3, we summarize
the features of the theoretical CMDs of stellar populations. In
Section 4, we present the binary fittings of the CMDs of star
clusters and show the properties of GCs. In Section 5, we compare
the results determined respectively by binary- and single-star
populations. Finally, in Sections 6 and 7, we give our discussion
and conclusions.


\section{Sample globular clusters and their photometry}
\label{sect:Obs} There are a lot of observational results for the
CMDs of globular clusters and some ones are presented recently
(e.g., \citealt{piotto02}, \citealt{kerber05}). We take some
observational results from a catalogue of the UBV HR diagrams of
globular clusters (\citealt{philip06}, hereafter Philip catalogue)
for this work. The reason is that these observational data show some
clear CMDs and seem reliable. In addition, photometry results have
less uncertainties than spectral results, and the photometry source
affects the final results slightly. Furthermore, we aims to try a
binary way to study the CMDs rather than to obtain very accurate
properties of star clusters. Thus the selected CMDs are suitable for
this work. The Philip catalogue is of 65 color-magnitude diagrams of
star clusters, which are compiled from the literature. Some star
clusters in the catalogue are shown a few CMDs derived from
different sources. When choosing the CMDs used in the work, some
ones with better CMD shapes are taken. As a result, we choose twelve
globular clusters from the catalogue. The twelve globular clusters
are NGC104, NGC2419, NGC4147, NGC4372, NGC5272, NGC5897, M5,
NGC6205, M10, NGC6352, NGC6397, and NGC6809. Note that all sample
clusters contain more than 100 observed stars. This possibly makes
the results more reliable.

\section{Theoretical colour-magnitude diagrams}
Theoretical CMDs are basic ingredients of the work. This work uses
two kinds of theoretical CMDs, i.e., the CMDs of bsSSPs and ssSSPs.
Both the two kinds of theoretical CMDs are built on the basis of an
isochrone database of stellar population studies (\citealt{li08}).
For convenience, we take a Salpeter shape (\citealt{salpeter55}) for
the initial mass function of theoretical stellar populations. When
modeling binary-star stellar populations, a fraction of 50\% is
taken for the binaries that have orbital periods less than 100 yr
(typical value of Milky Way Galaxy), and the lower and upper mass
limits are set to be 0.1 and 100 mass of solar, respectively. The
stellar evolution is calculated by a rapid star evolution code of
\cite{hurley02} (hereafter Hurley code). In each bsSSP, binary
interactions such as mass transfer, mass accretion, common-envelope
evolution, collisions, supernova kicks, angular momentum loss
mechanism, and tidal interactions are considered. Some default
parameters of the Hurley code, i.e., 0.5, 1.5, 1.0, 0.0, 0.001, 3.0,
190.0, 0.5, and 0.5, are taken for wind velocity factor ($\beta_{\rm
w}$), Bondi-Hoyle wind accretion faction ($\alpha_{\rm w}$), wind
accretion efficiency factor ($\mu_{\rm w}$), binary enhanced mass
loss parameter ($B_{\rm w}$), fraction of accreted material retained
in supernova eruption ($\epsilon$), common-envelope efficiency
($\alpha_{\rm CE}$), dispersion in the Maxwellian distribution for
the supernovae kick speed ($\sigma_{\rm k}$), Reimers coefficient
for mass loss ($\eta$), and binding energy factor ($\lambda$),
respectively. These values are taken because they have been tested
by \cite{hurley02} and seem more reliable than other ones. One can
refer to the paper of \cite{hurley02} for more details. Therefore,
many of these free parameters remain large uncertain, and they are
needed to be studied detailedly in the future. Accordingly, this
work is limited by the isochrone database, although it is a
convenient choice. In addition, the BaSeL 2.2 photometry library
(\citealt{lejeune98}) is used when transforming the isochrone
database into the CMDs of bsSSPs and ssSSPs. \label{sect:data}

\section{Binary fitting of colour-magnitude diagrams of globular clusters}
\label{sect:analysis} Basing on the theoretical CMDs of bsSSPs, we
fit the CMDs of our sample clusters and derive some basic parameters
of these clusters in this section. In order to get reliable results,
big ranges are taken for the stellar ages and metallicities of
theoretical populations when fitting the observational CMDs of star
clusters. In detail, the stellar age and metallicity ranges are
taken as 0.1 -- 15\,Gyr and 0.0001 -- 0.03, respectively. These are
just the default ranges of the isochrone database used by this work.
Although the largest metallicity of the theoretical populations is
only 0.03, it is enough for studying the CMDs of most globular
clusters, as globular clusters are usually metal-poor ($Z \leq$
0.03). Because the observational data of stars with high
luminosities have less observational uncertainties, a
magnitude-weighted method is taken in this work. This can possibly
enhance the reliability of the fitting results. For clearly, we show
the comparison of the observational and best-fit theoretical CMDs of
12 GCs in Figs. 1, 2 and 3. The best-fit CMDs are found by comparing
the distribution of stars in theoretical CMDs to that in
observational CMDs. The advantage of such a method is that both the
shape and luminosity function are compared at the same time.
\begin{figure}
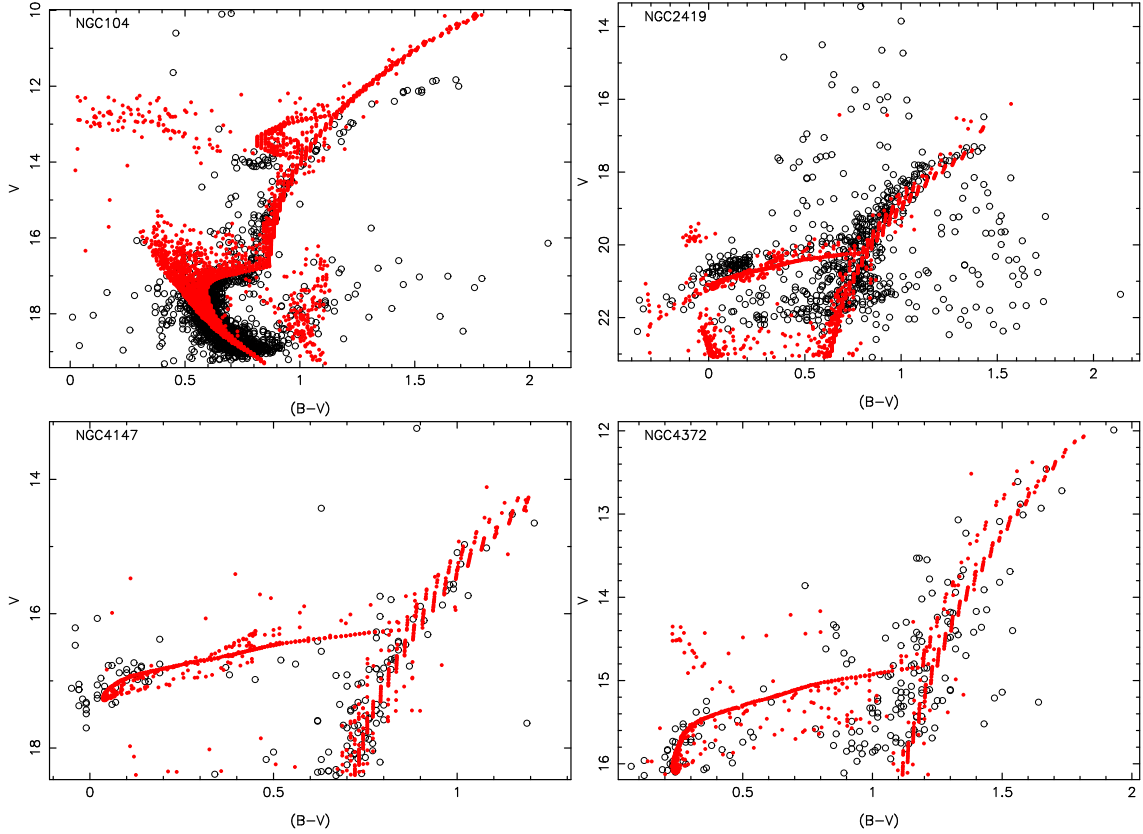


\includegraphics[angle=-90,width=75mm]{ms256fig1.ps}
\includegraphics[angle=-90,width=75mm]{ms256fig2.ps}
\includegraphics[angle=-90,width=75mm]{ms256fig3.ps}
\includegraphics[angle=-90,width=75mm]{ms256fig4.ps}

\caption{Comparison of observational (black circles) and best fit
theoretical (red points) CMDs, for globular clusters NGC104,
NGC2419, NGC4147, and MGC4372, respectively. The theoretical CMDs
are calculated from binary-star populations with 50\% binaries and
each star is assumed to be discriminable. In the field shown in the
figure, the observational and theoretical CMDs contain the same
stars, but a few red points contain less than 1 star.}
   \label{Fig1}
\end{figure}

\begin{figure}
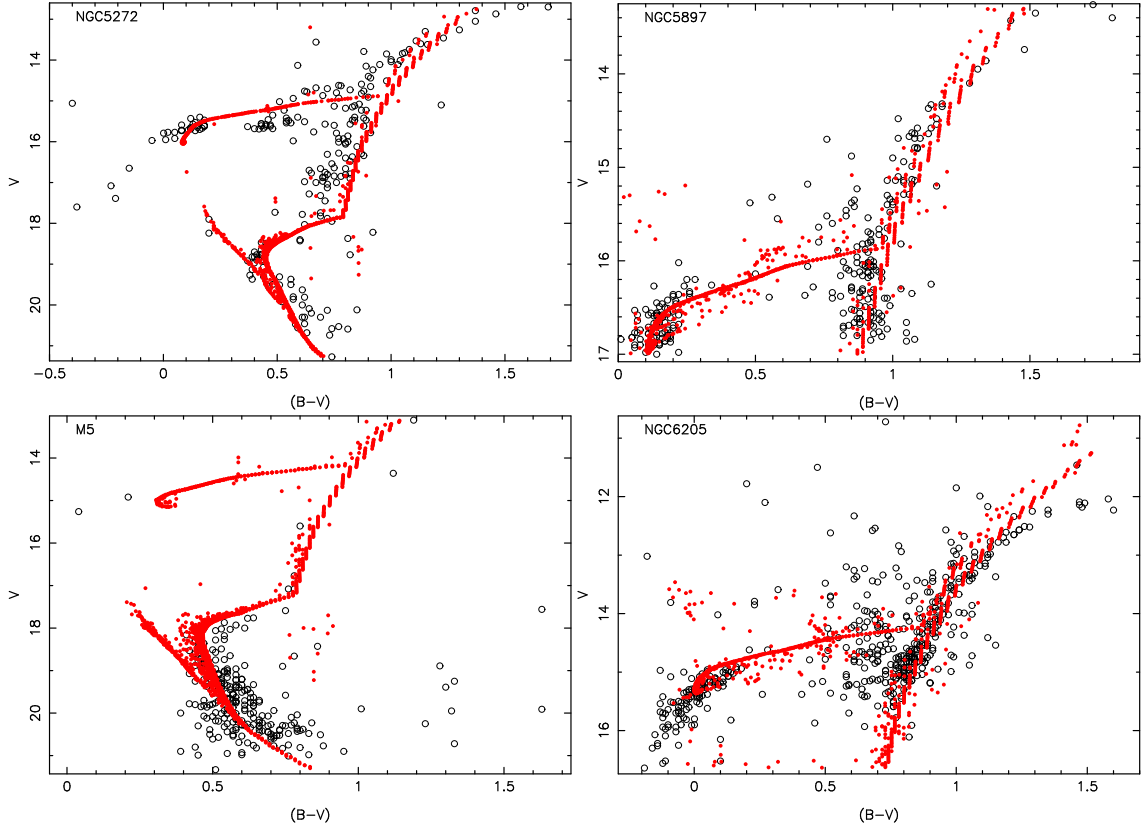


\includegraphics[angle=-90,width=75mm]{ms256fig5.ps}
\includegraphics[angle=-90,width=75mm]{ms256fig6.ps}
\includegraphics[angle=-90,width=75mm]{ms256fig7.ps}
\includegraphics[angle=-90,width=75mm]{ms256fig8.ps}

\caption{Similar to Fig.1, but for NGC5272, NGC5897, M5, and
NGC6205.}
   \label{Fig2}
\end{figure}

\begin{figure}
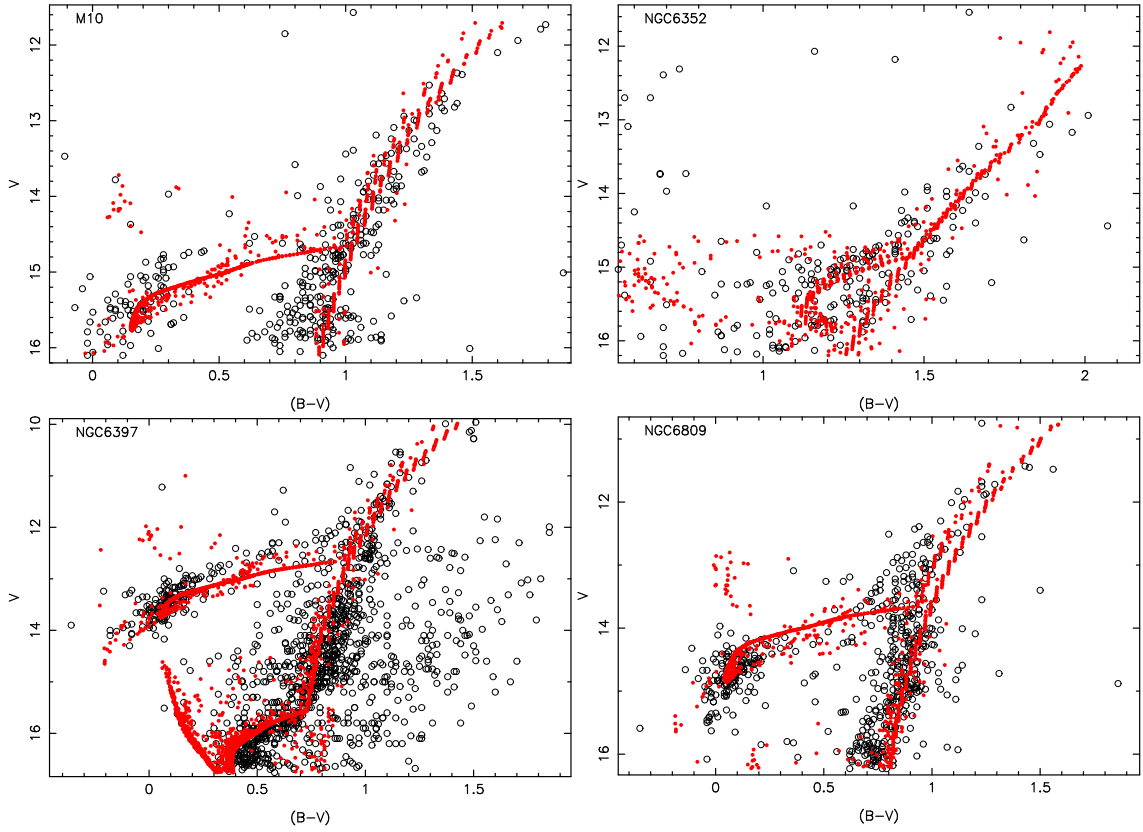


\includegraphics[angle=-90,width=75mm]{ms256fig9.ps}
\includegraphics[angle=-90,width=75mm]{ms256fig10.ps}
\includegraphics[angle=-90,width=75mm]{ms256fig11.ps}
\includegraphics[angle=-90,width=75mm]{ms256fig12.ps}

\caption{Similar to Fig.1, but for M10, NGC6352, NGC6397, and
NGC6809.}
   \label{Fig3}
\end{figure}

From the three figures shown we can see that the main shapes of the
CMDs of star clusters are well reproduced. Therefore, the the
properties of star clusters derived from the CMD fitting are
reliable. In detail, the stellar ages, metallcities, distances, and
reddenings determined by comparing observational CMDs to synthetic
CMDs are listed in Table 1. It shows that these clusters have a
minimum age of 4\,Gyr and a maximum age of 13\,Gyr, with an average
of 11.42\,Gyr. The lowest and highest metallicities ($Z$) of these
globular clusters are 0.0001 and 0.01, respectively, which has an
average of 0.00145. As a whole, the results show old ages and poor
metallicities for globular clusters. This is in agreement with most
previous studies.
\begin{table}[h!!!]

\small \centering

\begin{minipage}[]{120mm}

\caption[]{Best fit properties (stellar age, metallicity, distance
modulus, and reddening) of 12 globular clusters. The results are
obtained by comparing the observed CMDs to the CMDs of binary-star
stellar populations, and finding the best fit results. Note that a
magnitude-weighted method is used in the fitting.}\label{Table
1}\end{minipage}
\tabcolsep 4mm
 \begin{tabular}{ccccc}
  \hline\noalign{\smallskip}
Cluster Name &Age      & Metallicity ($Z$)&(m - M)$_V$ &$E(B-V)$\\
             &[Gyr]    &                  &[mag]            &[mag] \\
  \hline\noalign{\smallskip}
NGC104       &11.3     &0.0040            &13.4             &0.10  \\
NGC2419      &9.7      &0.0001            &20.8             &0.00  \\
NGC4147      &10.6     &0.0001            &16.8             &0.06  \\
NGC4372      &14.3     &0.0010            &15.4             &0.30  \\
NGC5272      &12.5     &0.0001            &15.4             &0.16  \\
NGC5897      &11.9     &0.0001            &16.4             &0.16  \\
M5           &9.5      &0.0003            &14.8             &0.12  \\
NGC6205      &12.4     &0.0003            &14.8             &0.04  \\
M10          &11.6     &0.0010            &15.2             &0.20  \\
NGC6352      &8.4      &0.0100            &15.6             &0.24  \\
NGC6397      &11.4     &0.0001            &13.2             &0.10  \\
NGC6809      &13.4     &0.0003            &14.2             &0.12  \\
  \noalign{\smallskip}\hline
\end{tabular}
\end{table}

\section{Comparison of different results}
Some bsSSP models are used for studying the CMDs of globular
clusters in this work, but most other works use ssSSP models. In
order to investigate the differences between the CMD studying
results derived from bsSSP and ssSSP models, we compare the
observational CMDs of two globular clusters (M5 and NGC6397) with
those of theoretical bsSSPs and ssSSPs. The two clusters are chosen
for the study because their CMDs are intact. In Fig. 4, the detailed
comparison of the observational CMDs with the best-fit theoretical
CMDs of bsSSPs and ssSSPs is shown. We see that the CMDs of best-fit
bsSSPs cover larger ranges in the colour versus magnitude field
compared to the best-fit ssSSPs, and the CMDs of bsSSPs are actually
closer to the observational CMDs of the two globular clusters. When
comparing the properties determined by bsSSPs and ssSSPs, they are
shown to be different. In detail, the best bsSSP fit age,
metallicity, distance modulus in $V$ band, and reddening $E(B-V)$ of
M5 are 9.5\,Gyr, 0.0003, 14.8\,mag, and 0.12\,mag, while the ssSSP
fit results are 12.6\,Gyr, 0.0001, 14.75\,mag, and 0.15\,mag,
respectively. The other star cluster, NGC6397, is shown some bsSSP
fit properties of 11.4\,Gyr (age), 0.0001 (metallicity $Z$),
13.2\,mag (distance modulus), 0.1\,mag (reddening), with ssSSP
results of 13.6\,Gyr, 0.0001, 13.0\,mag, 0.15\,mag, respectively.
Because the same method was used for the bsSSP and ssSSP fitting,
the differences result from the stellar population models.


\begin{figure}

\includegraphics[angle=0,width=75mm]{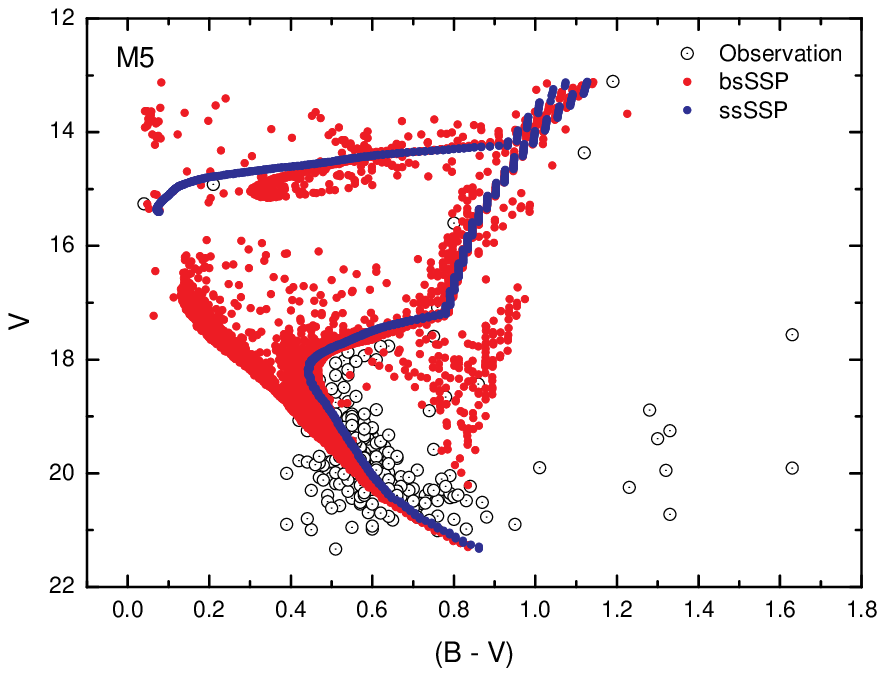}
\includegraphics[angle=0,width=75mm]{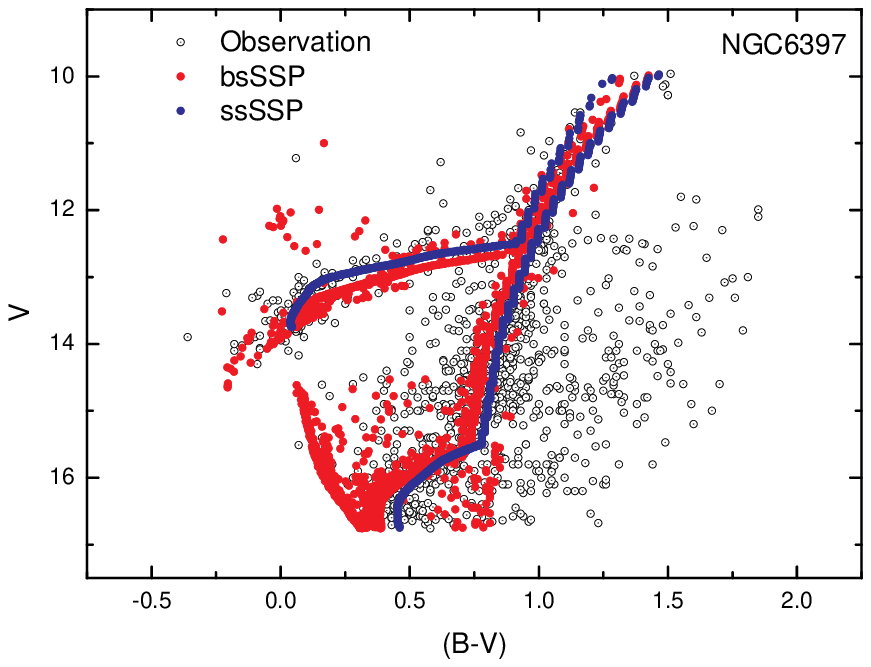}

\caption{Comparison of the observational CMDs (black) of star
clusters M5 and NGC6397 to the best fit bsSSP (red) and ssSSP (blue)
theoretical CMDs. A bsSSP take 50\% binaries for its population
stars and an ssSSP assumes that all population stars are single
stars.}
   \label{Fig4}
\end{figure}

\section{Discussion}
\label{sect:discussion} Binary stars can explain the observational
CMDs of globular clusters as a whole, but it is clear that there are
some differences between the observational and theoretical CMDs.
This possibly results from the following reasons: First, there are
some limitations in our theoretical stellar populations. For
example, the basic inputs (e.g., binary fraction, initial mass
function) of the bsSSP model remain some uncertainties and the
metallicity intervals are not enough small. Note that the binary
fractions in GCs are possibly lower than those ($\sim$ 50\%) in in
the local fields, open clusters, and star-forming regions, because
frequent dynamical interactions together with binary evolution
processes conspire to destroy binaries in GCs effectively. Second,
it seems impossible to discriminate every star of star clusters, but
each star is assumed to be distinguishable in our work as there is
no reliable instruction about how stars in a cluster can be
distinguished. This effect is obvious in the low luminosity part of
the main sequences of CMDs. Third, it seems that there are some
uncertainties in the observational CMDs. In addition, the effect of
superposition in CMDs has not been taken into account in this work,
this also affect morphology considerably, especially in crowded
regions such as the cluster center.

\section{Conclusions}
\label{sect:conclusion} This paper presents a binary-star study for
the colour-magnitude diagrams of 12 globular clusters, in which each
star is assumed to be distinguishable. It shows that the CMDs of
star clusters can be explained well via binary-star populations,
although some differences are shown in the comparisons of
observational and theoretical CMDs. As a result of the study, some
basic properties, i.e., stellar age, metallicity, distance modulus,
and colour excess, are determined for the 12 sample clusters. The
sample clusters are shown to be old (Age $\geq$ 8.4\,Gyr) and metal
poor ($Z \leq$ 0.01). When we compare the differences between the
cluster properties determined respectively by binary- and
single-star stellar populations, it shows that the two kinds of
population models can give different results for star clusters.
Because binary stars are common, the study suggests to use binary
method to study the CMDs of star clusters. It also suggests to model
the stellar populations of star clusters via binary stars.

\normalem
\begin{acknowledgements}
We thank the referee for comments and suggestions that helped
improve the paper. We are also thank Profs. Gang Zhao and Zhanwen
Han for useful discussions. This work was supported by Chinese
National Science Foundation (Grant No. 10963001), Yunnan Science
Foundation (2009CD093), and the Scientific Research Foundation of
Dali University (DYKF2009 No. 1).
\end{acknowledgements}

\label{lastpage}

\end{document}